\documentclass[sigconf]{acmart}
\setcopyright{none}
\copyrightyear{2024}
\acmYear{2024}
\acmDOI{XXXXXXX.XXXXXXX}

\acmConference[In2Writing '24]{the Third Workshop on Intelligent and Interactive Writing Assistants}{May 11}{Honolulu, HI}
\usepackage{colortbl}
\usepackage{graphicx}
\usepackage{enumitem} % added by Max for RQ numbering

\newcommand{\killpunct}[1]{}
 
\begin{document}
\settopmatter{printacmref=false}
\setcopyright{none}
\renewcommand\footnotetextcopyrightpermission[1]{}
\pagestyle{plain}

\title{The Dearth of the Author in AI-Supported Writing}

\author{Max Kreminski}
\affiliation{
\institution{Santa Clara University}
\streetaddress{500 El Camino Real}
\city{Santa Clara}
\state{California}
\country{USA}
}
\email{mkreminski@scu.edu}

%\renewcommand{\shortauthors}{Anderson, et al.}

\begin{abstract}
We diagnose and briefly discuss the \emph{dearth of the author}: a condition that arises when AI-based creativity support tools for writing allow users to produce large amounts of text without making a commensurate number of creative decisions, resulting in output that is sparse in expressive intent. We argue that the dearth of the author helps to explain a number of recurring difficulties and anxieties around AI-based writing support tools, but that it also suggests an ambitious new goal for AI-based CSTs.
\end{abstract}


%\ccsdesc[500]{Human-centered computing~Empirical studies in HCI}
%\ccsdesc[100]{Computing methodologies~Natural language processing}


\iffalse
\received{20 February 2007}
\received[revised]{12 March 2009}
\received[accepted]{5 June 2009}
\fi

\maketitle

\section{The Dearth of the Author}
What's the point of writing? Sometimes we write to learn, or write to think, or write for solitary enjoyment---but when the process of writing results in a piece of writing that's meant to be shared and read by others, what is this piece of writing \emph{for}? We assert that in most cases, a piece of writing is meant primarily to convey its author's \emph{expressive intent}: to communicate information, invoke feelings, or otherwise express to the reader a set of ideas and sentiments that the author intends to share. Indeed, this assumption forms the basis of the \emph{expressive communication} framework~\cite{ExpressiveCommunication} that has been used to evaluate AI-based creativity support tools (CSTs)~\cite{ShneidermanCSTs,CSTReviewFrich,CSTReviewChung} in the past.

The process of writing can thus be viewed as a process of \emph{decision-making} by the author. Authors of fiction must decide on character names and personalities and appearances, on details of setting and background, on the balance between action and introspection, on moment-to-moment emotional tone; authors of argumentative essays must decide on the points they want to argue, on the evidence they want to use in support of each point, on the order in which to introduce this evidence; authors in general must decide on what words they want to use in what places to get their ideas across. A piece of writing that contains fifteen hundred words can be thought of as the result of \emph{at least} fifteen hundred decisions: perhaps a bit fewer when some of these words are ``forced choices'' that could not be exchanged for another due to rules of grammar, but usually many more due to the need for decisions \emph{beyond} simple word choices in composing any useful or interesting piece.

Historically, an author could not compose a substantive piece of writing without making the vast majority of these decisions directly.\footnote{Though some words might be drawn verbatim from quotations, and in rare cases some might be chosen via mechanistic, aleatoric, or automatic writing processes, such as those adopted by the Oulipo movement~\cite{Oulipo}.} Thus the whole of every piece of writing could be taken as reflecting its author's intent: the relationship between the length of a written piece and the number of decisions that its author had to make was relatively fixed, and an author could not lengthen a piece of writing without deciding what specific words to add.

But we suddenly find that this is no longer the case. In particular, when authors use AI-based CSTs to expand a small amount of input text (such as a one-sentence instruction) to a large amount of output text (such as a complete written story or essay), they \emph{delegate} many of the creative decisions involved in producing the larger output to the CST---resulting in a piece of writing with an unusually low ratio of human decision-making to output length. In other words, the expressive intent of the author is underspecified relative to the amount of text that is generated, and the resulting piece of writing is unusually sparse in terms of expressive intent per word.

We refer to this unusual situation as the \textbf{dearth of the author}: the na\"ively AI-augmented author is not absent or dead, but their intent is stretched so thinly over their writing that they may feel barely present. In lieu of authorial intent, creative decisions are made by the CST to which the author has delegated portions of the writing process; simple LLM-based CSTs like ChatGPT make these decisions by approximating highly probable choices that a certain set of raters might also score well~\cite{InstructGPT}, while other CSTs lean on knowledge encoded in handcrafted rulesets~\cite{LooseEnds}, in large human-constructed databases of common-sense knowledge~\cite{Metaphoria}, or in more specialized corpuses of text~\cite{Dramatron}. Across the full range of AI-based CSTs for writing~\cite{WritingCSTs}, the greater the discrepancy between the size of a minimum viable user input and the size of the output piece of writing, the more the na\"ive user's results tend to exhibit the dearth.

\section{Some Implications of the Dearth}
Despite its simplicity, the dearth of the author helps to explain many of the disparate anxieties and difficulties around AI-based writing assistants. We briefly discuss a few of its implications here.

\textbf{Homogenization of writing.} Several recent studies of AI-based CSTs have found either direct~\cite{PredictableWriting,PadmakumarHe,DoshiHauser,HomogenizationEffectsLLMs} or indirect~\cite{OpinionatedLanguageModels,NextPhraseSuggestions} evidence that these tools can exert a \emph{homogenizing} effect on the creative outputs produced by different users: in other words, different users of the same CST may produce more similar outputs than they would without the CST. Homogenization effects can be explained by authors' delegation of creative decision-making to a tool that makes similar creative decisions in similar usage scenarios, with greater degrees of homogenization resulting from tools that make a greater proportion of creative decisions directly. 

\textbf{Limited feelings of ownership.} Recent studies have also found that users of AI-based CSTs for writing tend to experience a limited sense of ownership of or responsibility for the outputs of their interaction with the CST~\cite{CoAuthor,DoNotPerceiveOwnership}. This is similarly explained by the delegation of creative decisions to the CST: delegating a greater proportion of creative decisions to the CST seems likely to result in a commensurately lower feeling of ownership toward the resulting text, as seen both in \citeauthor{DoNotPerceiveOwnership}~\cite{DoNotPerceiveOwnership} and in the stronger sense of ownership reported by users of a narrower CST~\cite{LooseEnds}.

\textbf{Rhetorical confusion.} The typical outputs of AI-supported creative processes are often referred to as ``soulless''. Simultaneously, users who build highly intentful creative processes around AI-based CSTs---sometimes considering and discarding many dozens of AI outputs before accepting one as complete~\cite{ThePromptArtists}---are perplexed by these assertions. The dearth of the author helps explain both phenomena. The term ``soullessness'' reflects the sparseness of intent in the outputs that are easiest to create with many AI-based tools, and that therefore dominate most non-enthusiasts' impressions; meanwhile, the high-intent nature of some AI artists' processes may not be immediately apparent to onlookers, because AI tools permit the creation of comparably sized outputs from much smaller specifications of intent.

\textbf{Greater impacts on inexperienced writers.} Experienced writers tend to have an ear for evocative language and a strong aversion to clich\'e, both of which are components of a sophisticated sense of \emph{taste} built up over many years of paying close attention to language. These writers are readily able to identify problems in AI-generated text~\cite{TTCW}, and are often unwilling to simply accept creative decisions made by the machine when these decisions conflict with their own sensibilities~\cite{Dramatron}. Novice writers, on the other hand, may tend to treat the machine as a creative authority~\cite{HomogenizationEffectsLLMs} and delegate a greater proportion of their creative decisions---a form of \emph{algorithmic loafing}~\cite{AlgorithmicLoafing} that is likely to result in a stronger sense of dearth.

\textbf{The value of underdetermination.} Different AI-based CSTs for writing vary widely in the nature of their output and their ability to make a large proportion of the decisions involved in writing on the user's behalf. When the outputs of a CST take the form of ``sparks''~\cite{Sparks}, plot outlines~\cite{LooseEnds}, questions~\cite{IntentElicitation}, or critiques~\cite{WithoutWritingForThem} rather than output-ready prose, the user cannot as easily delegate creative decisions about the integration of these elements into a complete piece of writing---leaving open a space of \emph{underdetermination}~\cite{Underdetermination}, or perhaps even \emph{creative struggle}~\cite{CreativeStruggle}, into which expressive intent must flow. 

\section{Reversing the Dearth}
The dearth of the author arises when the ratio of authorial intent to output text length is small. AI-based CSTs that make creative decisions on the author's behalf tend to decrease this ratio. But AI-based CSTs can also \emph{increase} this ratio by leading the author to make \emph{a greater number of meaningful creative decisions per unit of text produced}, drawing out unexpressed elements of the author's intent and provoking them to refine this intent further~\cite{IntentElicitation,ReflectiveCreators}. CSTs that do this well enough may even bring about an unexpected alternative condition: an \textbf{abundance of the author}, in which every word of a piece has been considered more carefully and from more different angles than the author could otherwise manage or afford. The question of how exactly to design CSTs to support this abundance---we contend---is one of the biggest and most potentially impactful research questions that faces the field of AI-based writing support today.

\begin{acks}
This work was supported by Hackworth Grant GR102981, ``Investigating Homogenization of Imagination by Generative AI Models'', from the Markkula Center for Applied Ethics. Some of the ideas discussed here were initially formulated in response to discussions held at Stochastic Labs during the summer of 2022; thanks to all who participated in these discussions, and in particular to Joel Simon and Kyle Steinfeld.
\end{acks}

\bibliographystyle{ACM-Reference-Format}
\bibliography{bibliography}

%\appendix

\end{document}